\documentclass[10pt,peerreview,final,finalsubmission,twocolumn]{IEEEtran}
\newtheorem{thm}{Theorem}[section]
\newtheorem{cor}[thm]{Corollary}
\newtheorem{lem}[thm]{Lemma}

\newtheorem{defn}[thm]{Definition}
\usepackage{cite}
\usepackage{graphicx}
\usepackage{amsmath}
\usepackage{caption}
\usepackage{color}
\usepackage{ifpdf}
\usepackage{mathrsfs}
\usepackage{lineno,hyperref}
\usepackage{amssymb}
\usepackage{epstopdf}
\allowdisplaybreaks    

\begin{document}

\title{Hybrid Fault diagnosis capability  analysis of Hypercubes under the PMC model and MM$^{*}$ model }

\author{Qiang Zhu,Lili Li,  Sanyang Liu, Xing Zhang
\IEEEcompsocitemizethanks{\IEEEcompsocthanksitem  The authors are with the School of Mathematics and Statistics, Xidian University, Xi'an, Shaanxi 710071, China.\protect\\
E-mail: zhuustcer@gmail.com,liusanyang@126.com}
\thanks{Supported by National Natural Science Foundation of China (Nos: 61672025, 61674108, 61373174) }}


\IEEEcompsoctitleabstractindextext{%
\begin{abstract}
 System level diagnosis is an important approach for the fault diagnosis of multiprocessor systems.
In system level diagnosis, diagnosability is an important measure of the diagnosis capability of interconnection networks. But as a measure, diagnosability can not reflect the diagnosis capability of multiprocessor systems to link faults which may occur in real circumstances. In this paper, we propose the definition of $h$-edge tolerable diagnosability to better measure the diagnosis capability of interconnection networks under hybrid fault circumstances.  The $h$-edge tolerable diagnosability of a multiprocessor system $G$ is the maximum number of faulty nodes that the system can guarantee to locate when the number of faulty edges does not exceed $h$,denoted by $t_h^{e}(G)$.
The PMC model and MM model  are  the two most widely studied diagnosis models for the system level diagnosis of multiprocessor systems. The hypercubes are the most well-known interconnection networks. In this paper,  the $h$-edge tolerable diagnosability of $n$-dimensional hypercube under the PMC model and MM$^{*}$   is determined as follows:  $t_h^{e}(Q_n)= n-h$, where $1\leq h<n$, $n\geq3$.
\end{abstract}

}

\maketitle

\IEEEdisplaynotcompsoctitleabstractindextext

\IEEEpeerreviewmaketitle

\section{Introduction}
Nowadays, some supercomputers have hundreds of thousands of processors. It's inevitable that some node or link faults may occur in such large systems.
System level diagnosis is an approach  for the  fault diagnosis of multiprocessor systems by means of mutual self-tests of processors in the system.
Distinct definitions of test and distinct assumptions on test result lead to distinct diagnosis models.  PMC model\cite{F.P.Preparata1967} (introduced by Preparata, Metze and Chien) is the most widely studied model in system level diagnosis. Under the PMC model, a test involves two processors: the tester and the testee. It's assumed that only adjacent nodes can test the status of each other. As to the test results, it is assumed that if the tester  is fault-free(resp. faulty), then the test result is reliable(resp. unreliable).
Another important model, first proposed by Malek and Maeng \cite{Maeng1981A}, is called the comparison diagnosis model (MM model).   It is well known and widely studied in recent years.   In this model, a test $t(u,v;w)$ involves 3 processors $u, v, w$ where $w$ is the common neighbor of $u,v$. $w$ is called the comparator of $u, v$, $u,v$ are called the compared vertices. Under the MM model, it is assumed that the test result $r(u,v;w) $  is reliable if and only if the comparator $w$  is fault-free. That is, if $w$ is fault-free,  $r(u,v;w)= 0$ if and only if both $u, v$ are fault-free. If $w$ is faulty, $r(u,v;w)$ may be 0 or 1 irrelevant to the status of $u, v$. In 1992, Sengupta and Dahbura \cite{Sengupta1992On} suggested a further modification of the MM model, called the MM$^*$ model, in which every node must compare every pair of its distinct neighbors.   MM$^{*}$ model is widely used in the diagnosis capability analysis of interconnection networks \cite{chang2005diagnosabilities,Chang2007,chiang2012diagnosability,Lee2011Diagnosability,
hsieh2011diagnosability,Hsieh2008Diagnosability,YangM2013Conditional,Yang2013Conditional,Zhang2015The,Han2015The,Wang2016The}.

The performance of a large multiprocessor system is much impacted by its underlying topology which can be modeled by a graph called its interconnection network when each processor is regarded as a vertex and each communication link is regarded as an edge.

In the past two decades, various interconnection networks have been proposed. To choose an appropriate one, the properties of these interconnection networks have to be explored. Among all the properties of an interconnection network, fault diagnosis capability is very important to its suitability as the underlying topology of a high performance fault tolerant computing system.
Diagnosability, conditional diagnosability, $g$-good neighbor conditional diagnosability and $h$-extra diagnosability are some parameters for measuring the diagnosis capability of interconnection networks.The diagnosability of a multiprocessor system is the maximum number of faulty nodes that the system can guarantee to locate. That is, suppose the diagnosability of a system $G$ is $t$, then  for all circumstances where the number of faulty nodes in $G$ does not exceed $t$, all these faulty nodes can be located in one step and there exists a circumstance with $t+1$ faulty vertices such that not all faulty nodes can be located in one step.
 As an important measure of the diagnosis capability, the diagnosability of various interconnection networks  have been studied \cite{fan1998diagnosability,fan2002diagnosability,wang1999diagnosability,lai2004diagnosability,zheng2005diagnosability,
hsieh2011diagnosability,chiang2012diagnosability,wang1994diagnosability,zhou2013fault}.

Diagnosability is a worst-case measure, in real circumstances many interconnection networks exhibits much higher fault diagnosis capability compared with their diagnosability\cite{Somani1996On}.  To better measure the diagnosis capability of such interconnection networks, some new parameters like conditional diagnosability, $g$-good neighbor diagnosability and $h$-extra diagnosability have been
proposed and studied.

For some large interconnection networks like hypercubes, all the neighbours of any node are faulty  at the same time is quite small. So  conditional diagnosability has been proposed by Lai et al.   in 2005 \cite{Lai2005Conditional} by restricting that any node in the multiprocessor system must have a fault-free neighbor. The conditional diagnosability of various interconnection networks have been investigated \cite{Lai2005Conditional,Hsu2009Conditional,Hong2012Strong,Hsieh2013The,Hsieh2013Strong,
Li2014Conditional,Lin2015The,Zhu2008On,Zhu2008Onc,Chang2012Conditional,YangM2013Conditional,
Yang2013Conditional,Chang2015Conditional}.
In 2012, Peng et al. proposed $g$-good neighbor conditional diagnosability as a generalization of the conditional diagnosability by assuming that any node in a multiprocessor system much have at least $g$-good neighbors\cite{Peng2012a}. The $g$-good conditional diagnosability of several interconnection networks have been explored.
  In 2015, Zhang et al.   \cite{Zhang2015The} proposed $g$-extra conditional diagnosability by assuming that when some nodes fail there are no components with less than $g+1$ vertices in the remaining system.   The $g$-extra diagnosability of several well-known interconnection networks such as hypercubes, fold hypercubes, arrangement graph and bubble-sort star graph have been studied \cite{Zhang2015The,Han2015The,Xu2016The,Wang2016The}.

Although these parameters may better measure the diagnosis capability of interconnection networks, they are all only applicable to the circumstance with no link faults in the system. But in real circumstance, both node and link faults may occur  when a network is put into use. It is significant to study the diagnosis capability of multiprocessor systems with both node and link faults.   In 1998, CL Yang and GM Masson studied the hybrid fault diagnosability of multiprocessor systems with unreliable communication links by introducing bounded number of incorrect test results \cite{yang1988hybrid}.    In 2000, D Wang discussed the strategies for determining  the diagnosability of hypercubes with arbitrary missing edges under both the PMC model and BGM model \cite{wang2000diagnosability}.   In 2007, S Zheng and S Zhou proved that under the PMC model and the BGM model, the diagnosability of star graphs with missing edges can be determined by its minimum  degree \cite{zheng2007diagnosability}.   In 2012, CF Chiang et al.   showed that under the MM$^*$ model the $n$-dimensional star graph preserves the strong local diagnosability when the number of faulty links does not exceed $n-3$ \cite{chiang2012diagnosability}.

To better adapt to the real circumstances where both node and link faults may occur,   we  introduce the $h$-edge tolerable diagnosability  to better measure the diagnosis capability of multiprocessor systems.   Then we investigate the $h$-edge tolerable diagnosability of hypercubes under the PMC model and MM$^*$ model.

The rest of this paper is organized as follows.   Section 2 introduces some definitions, notations and terminologies; In section 3, Preliminaries on the PMC and MM$^*$ model, the concept of $h$-edge tolerable diagnosability and some basic results about it are  presented.   In section 4, the $h$-edge tolerable diagnosability of the hypercube under the PMC model and MM$^*$ model are   explored.   Section 4 summarizes the results of this paper and gives some possible  future directions.

\section{Notations and Terminologies}
We follow \cite{Xu2013Theory} for terminologies and notations not defined here.   Let $G=(V(G),E(G))$ be a simple undirected graph, $V(G)$ and $E(G)$ are used to denote its vertex set and edge set respectively.   We use the unordered pair $(u,v)$  for an edge $e=\{u,v\}$.   If $(u,v)\in E(G)$, then $u$ and $v$ are adjacent and $u$, $v$ are incident to the edge $(u,v)$.   We use $E_G(u)$ (resp.   $N_G(u)$) to denote the set of edges incident (resp.   vertices adjacent) to $u$ in $G$.   The number of vertices in $N_G(u)$ is called the degree of $u$ in $G$, denoted by $deg_G(u)$.   The minimum degree of $G$ is the minimum over all degrees of the  vertices in $G$, denoted by $\delta(G)$.   If for any vertex $u\in G$, $deg_G(u)=k$, then the graph is said to be $k$-regular.   When $G$ is clear from the context, we use $E(u)$, $N(u)$ and $d(u)$ to respectively replace $E_G(u)$ and $N_G(u)$ and $deg_G(u)$ for simplification.   For any nonempty vertex set $S$ of $G$, we use $N_G(S)$ to denote the set of  vertices in $V(G)-S$ which are adjacent to at least one vertex in $S$, that is, $N_G(S)=\{u\in V(G)-S\; | \; \exists \; v \in S \mbox{ such that } (u,v)\in E(G)\}$.
 Let $C_G(S) = N_G(S)\cup S$.
For two graphs $H$ and $G$, we say $H$ is a subgraph of $G$ if $V(H)\subseteq V(G)$ and $E(H)\subseteq E(G)$, denoted by $H\subseteq G$.    The vertex boundary number of $H$ is the number of vertices in $N_G(V(H))$, denoted by $b_{v}(H, G)$.   By definition, it's obvious that subgraphs with the same vertex set have the same boundary number.   The minimum $m$-boundary number of $G$ is defined as the minimum boundary number of all its subgraphs with order $m$, denoted by $\delta_{v}(m;G)$.   That is, $\delta_{v}(m;G)= \{b_{v}(H, G)\;|\; H\subseteq G \mbox{ and } |V(H)|=m \}$.

Given a graph $G= (V, E)$, the symmetric difference of any two vertex subsets $A$ and $B$ is the set of elements in exactly one of $A, B$,
denoted by $A\triangle B$.   That is, $A\triangle B=(A\cup B)-(A\cap B)$.   The complement of $A$ in $G$ is $V(G)-A$, denoted by $\overline{A}$.


\section{Preliminaries and Basics of the $h$-edge tolerable diagnosability}

\subsection{Preliminaries on The PMC model}
In 1967, Preparata, Metze and Chien\cite{F.P.Preparata1967} proposed a diagnosis model  called the PMC model which is the most famous and most widely studied model in the system level diagnosis of multiprocessor systems. Under this model, it is assumed that only node faults can occur and all node faults are permanent. Under this model, a test involves two adjacent processors： a tester and a testee. It is assumed that the status of the testee can always be detected by a fault-free tester.  A test $(u,v)$ where $u$ is the tester and $v$ is the testee can be represented by an ordered pair $(u, v)$. The test result of $(u,v)$ is denoted by $r(u, v)$. $r(u,v)=0$ if $u$ evaluates $v$ as fault-free and $r(u,v)=1$ if $u$ evaluates $v$ as faulty. Under the PMC model, it is assumed that test result $r(u,v)$ is reliable if and only if  the tester $u$ is fault-free. That is, if $u$ is fault-free, then $r(u,v)=0$ means that $v$ is fault-free and $r(u,v)=1$ means that $v$ is faulty; If the tester $u$ is faulty, then the test result $r(u,v)$ may be 0 or 1 whether
  $v$ is faulty or fault-free.

\subsection{Preliminaries on The MM model}
Proposed by Malek and Maeng \cite{Maeng1981A},  the comparison diagnosis model (MM model) is one of the most popular diagnosis models.   In the MM model,   A test  $(u,v)_{w}$ is done by sending the same task from a node  $w$(the comparator) to a pair of its distinct neighbors: $u$ and $v$, then comparing their results.   The test result of $(u, v)_{w}$ ($r((u, v)_{w})$) is 0 (resp.   1) if $w$ evaluates that the returning results of $u$ and $v$ are consistent (resp.   not consistent).   Under the MM model, it is assumed that if the comparator $w$ is fault-free, $r((u, v)_{w})=0 $  if $u$ and $v$ are both fault-free and $r((u, v)_{w})=1$ if at least one of $u$, $v$ is faulty.   If the comparator $w$ is faulty, then the test result may be 0 or 1 irrelevant to the status of $u$ and $v$.


\subsection{Basics about the PMC model and MM model }
Given a multiprocessor system, the set of all test results is called a syndrome of the multiprocessor system. Under the PMC model, a fault set $F$ is said to be consistent with a syndrome $\sigma$  if $\sigma$ can be aroused by the circumstance that all nodes in $F$ are faulty and all nodes not in $F$ are fault-free.
 Since the test result of a faulty tester(resp.comparator ) is unreliable under the PMC model(resp. MM model), a fault set $F$ can be consistent with many syndromes, the set of all syndromes consistent with $F$f is denoted by $\sigma(F)$.
Two faulty sets $F_1, F_2$ are distinguishable if and only if $\sigma(F_1)\cap \sigma(F_2)=\emptyset$. Otherwise, they are indistinguishable. Since the test result of a faulty tester(resp. comparator) under the PMC model(resp. MM model) is unreliable, $V(G)$ is consistent with any syndrome of $G$ under both the PMC model and MM model. Thus to locate faulty vertices, people often suppose there exists an upper bounder for the number of faulty vertices.
A multiprocessor system $G$ is $t$-diagnosable if all the faulty vertices can be guaranteed to be located provided that the number of faulty vertices does not exceed $t$.  The diagnosability of $G$ is the maximum integer $t$ such that $G$ is $t$-diagnosable. The diagnosability of a multiprocessor system can measure its fault diagnosis capability.

In \cite{Dahbura1984} A. Dahbura et. al. characterize a pair of distinguishable faulty sets under the PMC model.
\begin{lem}\label{dispmc}
For a simple undirected graph $G=(V,E)$ and any two different sets $F_1, F_2\subset V$, $F_1$ and $F_2$ are distinguishable  under the PMC model if and only if there exists an edge between $V-F_1\cup F_2$ and $F_1\Delta F_2$.
\end{lem}

Under the MM$^*$ model, Sengupta et al.   \cite{Sengupta1992On} provide a necessary and sufficient condition to check whether a pair of faulty sets is distinguishable.

\begin{lem}\label{dis-MM} \cite{Sengupta1992On} For a simple undirected graph $G=(V,E)$ and any two different sets $F_v^1, F_v^2\subset V$, $F_v^1$ and $F_v^2$ are distinguishable  under the MM$^*$ model if and only if at least one of the following conditions is satisfied:

(1) There exist three vertices $u\in (F_v^1\bigtriangleup F_v^2)$ and $v,w\in \overline{F_v^1\cup F_v^2}$ such that $uw,vw\in E(G)$.

(2) There exist three vertices $u,v\in (F_v^1-F_v^2)$ and $w\in \overline{F_v^1\cup F_v^2}$ such that $uw,vw\in E(G)$.

(3) There exist three vertices $u,v\in (F_v^2-F_v^1)$ and $w\in \overline{F_v^1\cup F_v^2}$ such that $uw,vw\in E(G)$.
\end{lem}

By Lemma \ref{dispmc} and Lemma \ref{dis-MM}, we have the following corollary:
\begin{cor}\label{pmcmm}
For a simple undirected graph $G=(V,E)$ and any two different sets $F_v^1, F_v^2\subset V$,  $F_v^1$ and $F_v^2$ are distinguishable  under the PMC model if they are distinguishable under the MM$^*$ model
\end{cor}

\subsection{The $h$-edge tolerable diagnosability}

To better adapt to the real circumstances that link faults may happen, we introduce the $h$-edge tolerable diagnosability of multiprocessor systems to better measure their diagnosis capability.

Given a multiprocessor system $G(V, E)$, we want to evaluate the diagnosis capability of the system when some edges are faulty. To do this, suppose $F_e$ is the set of faulty edges in the system.Then given $F_e$ as the set of faulty edges, under the MM$^*$ model two faulty vertex sets $F_v^1$ and $F_v^2$ are distinguishable if and only if they are distinguishable in $G-F_e$.

According to Lemma \ref{dispmc},
the following corollary can be obtained:
\begin{cor}\label{indispmc}
Given a multiprocessor system $G(V,E)$, let $F_e$ be the set of faulty edges in $G$. Then two faulty vertex sets $F_v^1$ and $F_v^2$ are indistinguishable in $G-F_e$ under the PMC  model if and only if in the graph $G- F_e$ any vertex in $V- F_v^1\cup F_v^2$ has no neighbor in $F_v^1\Delta F_v^2$.
 \end{cor}

According to Lemma \ref{dis-MM},
the following corollary can be obtained:

\begin{cor}\label{indismm}
Given a multiprocessor system $G(V,E)$, let $F_e$ be the set of faulty edges in $G$. Then two faulty vertex sets $F_v^1$ and $F_v^2$ are indistinguishable in $G-F_e$ under the MM$^*$  model if and only if in the graph $G- F_e$ any vertex in $V- F_v^1\cup F_v^2$ has at most one neighbor in $F_v^1- F_v^2$ or $F_v^2 -F_v^1$ and if it has a neighbor in $F_v^1\Delta F_v^2$,  then it doesn't has any neighbor in $V-F_v^1\cup F_v^2$.
 \end{cor}

%

%

Remember that the diagnosability of  a multiprocessor system $G$ under the MM model is the maximum integer $t$ such that $G$ is $t$-diagnosable.   To better measure the diagnosis capability of multiprocessor systems with link faults,  we propose the definition of $h$-edge tolerable diagnosability as a generalization of diagnosability under the MM model.

\begin{defn}\label{h-edge}
Given a diagnosis model and a multiprocessor system $G$, $G$ is $h$-edge tolerable $t$-diagnosable under the diagnosis model  if any pair of faulty vertex sets $F_v^1$ and $F_v^2$ are distinguishable in $G-F_e$  provided that $|F_v^1|, |F_v^2|\le t$ and $|F_e|\le h$ where  $F_e\subset E(G)$ is any edge  subset of $G$ with not more than $h$ edges;
 The $h$-edge tolerable diagnosability of  $G$, denoted as $t_{h}^{e}(G)$, is the maximum integer $t$ such that $G$ is $h$-edge tolerable $t$-diagnosable.
\end{defn}

The traditional diagnosability of a multiprocessor system $G$ under the MM$^*$ model can be viewed as the 0-edge tolerable diagnosability of $G$. Thus $h$-edge tolerable diagnosability is a generalization of the traditional diagnosability and can better measure the diagnosis capability of interconnection networks.

\section{the $h$-edge tolerable diagnosability of hypercubes under the PMC and MM$^*$ model}
\subsection{Preliminaries on the hypercubes }
Hypercubes are the most famous and most widely studied interconnection networks.   An $n$-dimensional hypercube $Q_n$ can be modeled as a graph where each vertex is labelled with an $n$-bit binary string.   Any pair of distinct vertices in $Q_n$ are adjacent if and only if their labels differ in exactly one bit position.   In other words, $u=u_1 u_2\cdot\cdot\cdot u_{n}$ and $v=v_1v_2\cdot\cdot\cdot v_{n}$ are adjacent if and only if there exists a positive integer $i\in\{1,2,\cdots,n\}$ such that $u_i\neq v_i$ and $u_j= v_j$, for each $j\in\{1,2,\cdot\cdot\cdot,n\}/\{i\}$.
In this case, we call $v$ an $i$-th neighbor of $u$, denoted by $u^i$.   Similarly $u$ can be denoted as $v^i$.   The edge $(u, u^i)$ is called the $i$-th incident edge of $u$.
 Clearly, $Q_ n$ is a $n$-regular graph consisting of $2^n$ vertices and $n\cdot 2^{n-1}$ edges, the minimum length of all cycles of $Q_n$ equals $4$ for $n \geq 2$.

The properties of $Q_n$ have been extensively studied.   In \cite{Qiang2006On} Zhu et al.   showed that any two different vertices in $Q_n$ can either have two common neighbors or none.

\begin{lem}\label{neighbours}\cite{Qiang2006On}
Any two distinct vertices in $V(Q_n)$ have exactly two common neighbours for $n\geq3$ if they have any.
\end{lem}


The results of the $m$-minimum boundary number of $Q_n$ is useful in this paper.

\begin{lem} \label{minimum}\cite{Yang2006Minimum}
\begin{displaymath}
\delta_{v}(m;Q_{n}) =\left\{\begin{array}{ll}
 - \frac{m^2}{2} +(n-\frac{1}{2})m+ 1 ,  & \textrm{$ 1\leq m \leq n+1 $}\\
-\frac{m^2}{2}+(2n-\frac{3}{2})m-n^2+2 .    & \textrm{$ n+2\leq m \leq 2n $}\\
\end{array} \right.
\end{displaymath}

\end{lem}

By Lemma \ref{minimum}, the following corollary can be obtained.
\begin{cor}\label{increase}
For any two integers $m_1,m_2$ with $1\le m_1 \le m_2 \leq 2n-2 $, $\delta_{v}(m_2;Q_{n})\geq \delta_{v}(m_1;Q_{n})-1$.
\end{cor}

\subsection{Main Results}
In this section, we will study the $h$-edge tolerable diagnosability of $Q_n$ under both the PMC model and the  MM$^*$model.

\begin{lem}\label{upper}For an $n$-dimensional hypercube $Q_n$ with  $n\geq3$, $t_h^{e}(Q_n)\leq n -h$ where $1\leq h\leq n$ under both the PMC model and the  MM$^*$model.
\end{lem}

$\mathbf{Proof:}$ To prove this Lemma, we only need to construct two distinct  faulty vertex sets $F_v^1$,  $F_v^2$  and a faulty edge set $F_e$ satisfying $|F_e|\le h$ and $|F_v^1|,|F_v^2|\le n-h+1$ and $F_v^1$,  $F_v^2$  are indistinguishable in $Q_n -F_e$.

  Let $u=0^n$, $F_e=\{(u,u^1), (u, u^2) \cdots (u,u^h)\} $, $F_v^1=\{u^{(h+1)},u^{(h+2)},\cdots,u^{n}\}$,

  $F_v^2=\{u,u^{(h+1)},u^{(h+2)},\cdots,u^{n}\}$.
Since in $Q_n -F_e$ there is no edge between $V- F_v^1\bigcup F_v^2$ and $F_v^1\Delta F_v^2$, $F_v^1$ and $F_v^2$ are indistinguishable in $Q_n -F_e$ under both the PMC model and the MM$^*$ model according to Corollary  \ref{indispmc} and Corollary \ref{indismm}.   Thus  $t_h^{e}(Q_n)<\max\{|F_v^1|, |F_v^1|\}=  n -h+1$ under both  the PMC model and the MM$^*$ model.
The lemma holds.  \hfill\rule{1mm}{2mm}

Next, we show that the above upper bound of $t_h^{e}(Q_n)$ can be reached  under both models.

\begin{lem}\label{lower}
Under both the PMC model and the  MM$^*$model. $t_h^{e}(Q_n)\geq n-h$, where $1\leq h\leq n$, $n\geq5$.
\end{lem}

$\mathbf{Proof:}$ Suppose, by contradiction, that $t_h^{e}(Q_n)<n-h$ under both  models.   By Definition \ref{h-edge} and Corollary \ref{pmcmm}, there exists  $F_e,\subset E(Q_n)$ and  $F_v^1$, $F_v^2\subset V(Q_n)$ with $|F_e|\leq h$, $|F_v^1|, |F_v^2|\leq n-h$ such that  $F_v^1$  $F_v^2$ are indistinguishable in $Q_n-  F_e$ under the MM$^*$ model.    By Theorem \ref{dis-MM}, in the graph $Q_n- F_e$ any vertex in $V- F_v^1\cup F_v^2$ has at most one neighbor in $F_v^1- F_v^2$ or $F_v^2 -F_v^1$ and if it has a neighbor in $V-F_v^1\Delta F_v^2$  then it doesn't has any neighbor in $V-F_v^1\cup F_v^2$.
So the following proof is divided into two cases: 1) $N_{V-F_v^1\cup F_v^2}(F_v^1\triangle F_v^2)=\emptyset$. 2) $N_{V-F_v^1\cup F_v^2}(F_v^1\triangle F_v^2)\not=\emptyset$

Case 1). $N_{V-F_v^1\cup F_v^2}(F_v^1\triangle F_v^2)=\emptyset$

In this case, $N_{Q_n -F_e}(F_v^1\Delta F_v^2)\subset F_v^1\cap F_v^2$. So $|F_v^1|+|F_v^2|\ge |F_v^1\cup F_v^2|=|F_v^1\cap F_v^2|+|F_v^1\Delta F_v^2| \ge |N_{Q_n -F_e}(F_v^1\Delta F_v^2)||+ |F_v^1\Delta F_v^2|$. For simplicity, let $m= |F_v^1\triangle F_v^2|$.

Subcase 1.1) $m=1$.

Without loss of generality, suppose $F_v^1-F_v^2\neq\emptyset$, then $F_v^1\cap F_v^2=F_v^2$.   Let $\{u\}=F_v^1-F_v^2$ Since $N_{V-F_v^1\cup F_v^2}(F_v^1\triangle F_v^2)=\emptyset$, all the  end-vertices of $u$ in $Q_n -F_e$ are in $ F_v^2$.   So  $|F_v^1|\ge n-h+1$, a contradiction.

Subcase 1.2) $m= 2$.

It's obvious that $|N_{Q_n -F_e}(F_v^1\Delta F_v^2)|\ge 2n-2 -h$. So  $2n-2h\ge |F_v^1|+|F_v^2|\ge |F_v^1\cup F_v^2|
\ge |N_{Q_n -F_e}(F_v^1\Delta F_v^2)||+ |F_v^1\Delta F_v^2|\ge 2n-2-h+2= 2n-h$, a contradiction.

Subcase 1.2) $m\ge 3$.

Since $|F_v^1|,|F_v^2|\leq n-h$, $m\leq 2n-2h$ and $1\leq h\leq n$.   By Corollary \ref{increase}, it is easy to see that $|N_{Q_n}(F_v^1\Delta F_v^2)|\geq \delta_v(3;Q_n)-1=3n-6$.  Therefore,
 $2n-2h\ge  |F_v^1\cup F_v^2|
\ge |N_{Q_n -F_e}(F_v^1\Delta F_v^2)||+ |F_v^1\Delta F_v^2|\ge (3n-6-h)+3= 3n-h-3$, a contradiction to $n\ge 3, h\ge 1$.
%

Case 2) $N_{V-F_v^1\cup F_v^2}(F_v^1\triangle F_v^2)\not=\emptyset$.

Let $g= |N_{V-F_v^1\cup F_v^2}(F_v^1\triangle F_v^2)|$,


Let $u\in F_v^1\Delta F_v^2$, without loss of generality, suppose $u\in F_v^1 -F_v^2$. Since $N_{V-F_v^1\cup F_v^2}(F_v^1\triangle F_v^2)\neq\emptyset$, suppose there exists a vertex $v\in V-F_v^1\cup F_v^2$ such that $uv\in E(u)-F_e$.   Since the number of the faulty edges is not more than $h$, obviously, $d_{G^{'}}(v)\geq n-h$.   Since $F_v^1,F_v^2$ are indistinguishable, at most one neighbouring vertex of $v$ is in $F_v^2-F_v^1$ and the rest neighbouring vertices of $v$ in $Q_n -F_e$ are in $F_v^1\cap F_v^2$.   So $|F_v^1\cap F_v^2|\geq n-h-2$. Let $G^{'}  = Q_n -F_e$.    Let $S=N_{G'}(v)\cap(F_v^2-F_v^1)$ and $|S|=g$, then $g\leq1$.   In this case, we consider the following two subcases depending on the size of $g$.

Subcase 2.  1 $g=0$

Since $F_v^1,F_v^2$ are indistinguishable, then $N_{G'}(v)\subset F_v^1$, $|F_v^1|\geq|N_{G'}(v)|\geq n-h$ and  combining with $|F_v^1|\leq n-h$, so $|F_v^1|=|N_{G'}(v)|=n-h$, $F_e\subset E(v)$, $F_v^2\cap F_v^1=N_{G'}(v)-u$.

If $N_{V-F_v^1\cup F_v^2}(u)-v\neq\emptyset$, suppose  $w\in N_{V-F_v^1\cup F_v^2}(u)-v$ (see Fig.  2).   Since $F_v^1, F_v^2$ are indistinguishable, $w$ has at most one neighbor in $F_v^1 -F_v^2$ or $F_v^2 -F_v^1$ and no neighbor in $V- F_v^1\cup F_v^2$. Since $F_e \in N(v)$, $|N_{G'}(w)\cap (F_v^1\cap F_v^2)|\geq n-2$.   Therefore, combining with lemma \ref{neighbours}, for $n\geq4$, we have
\input{h-2.TpX}
\begin{align}
n-h\ge |F_v^1|=&|F_v^1-F_v^2|+|F_v^1\cap F_v^2|\nonumber\\
    \geq&1+|N_{F_v^1\cap F_v^2}(v)\cup N_{F_v^1\cap F_v^2}(w)|\nonumber\\
    \geq&1+|N_{F_v^1\cap F_v^2}(v)|+|N_{F_v^1\cap F_v^2}(w)|-1\nonumber\\
    \geq&1+n-h-1+n-2-1\nonumber
    \end{align}
a contradiction to $n\ge 4$.

If $N_{V-F_v^1\cup F_v^2}(u)-v=\emptyset$, since $F_v^2\cap F_v^1=N_{G'}(v)-u$, $u,v$ don't have common neighbours in $F_v^1\cap F_v^2$ because of the minimum length of all cycles of $Q_n$ equals 4 for $n\geq3$, then all neighbours of $u$ aren't in $F_v^1\cap F_v^2$, namely, ${N_{G'}(u)-v}\subset F_v^2-F_v^1$, so $|F_v^2-F_v^1|\geq|N_{G'}(u)-v|=n-1$.   It is easy to obtain that for $n\geq3$,
\begin{align}
n-h\ge |F_v^2|=&|F_v^1\cap F_v^2|+|F_v^2-F_v^1|\nonumber\\
 \geq&|N_{G'}(v)-u|+|N_{G'}(u)-v|\nonumber\\
 \geq&(n-h-1)+(n-1)\nonumber
\end{align}
A contradiction.

Subcase 2.  2 $g=1$

Let $x$ be a neighbouring vertex of $v$ in $F_v^2-F_v^1$, then $|F_v^1\cap F_v^2|\geq|N_{G'}(v)|-2\geq n-h-2$.   Combining with $|F_v^1|\leq n-h$, $|F_v^2|\leq n-h$, we can obtain that
$$|F_v^1- F_v^2|\leq2\;\; \text{ and }\;\;|F_v^2- F_v^1|\leq2$$
We will distinguish between the following two cases.

Subcase 2.  2.  1 $|F_v^1-F_v^2|=2$ or $|F_v^2-F_v^1|=2$

Without loss of generality, we assume that $|F_v^1-F_v^2|=2$.   Let $u,y\in F_v^1-F_v^2$, then $$|F_v^1|\geq|F_v^1\cap F_v^2|+|F_v^1-F_v^2|\geq n-h-2+2=n-h,$$
so $|F_v^1|=n-h$, $|F_v^1\cap F_v^2|=n-h-2$ and $F_e\subset E(v)$.   Observe that this situation is similar to above mentioned subcase 2.1 , with  similar arguments,  a contradiction can be obtained.

Subcase 2.  2.  2 $|F_v^1- F_v^2|=1$ and $|F_v^2- F_v^1|=1$

Let $\{u\}=F_v^1- F_v^2$, $\{w\}=F_v^2- F_v^1$.   Hence, $v$ is adjacent to $u$ and $w$ and $uv,vw$ are fault-free edges.   Since $|F_v^1\cap F_v^2|\geq n-h-2$, $|F_v^1|\leq n-h$ and $F_v^1,F_v^2$ are indistinguishable,  ${N_{G'}(v)-w}\subset F_v^1$, $d_{G'}(v)-1\leq|F_v^1|\leq n-h$, therefore, $|F_e\cap E(v)|\geq h-1$.

If there exists a vertex $x\in B$, such that $x\in N_{G'}(u)$, then the other end-vertices of all the fault-free edges incident to $x$ have at least $n-3$ in $F_v^1\cap F_v^2$ because of $F_v^1,F_v^2$ are indistinguishable.   By lemma \ref{neighbours}, $x,v$ have exactly two common neighbours for $n\geq3$ if they have any.   Therefore, for $n\geq5$,
 \begin{align}
|F_v^1|=&|F_v^1\cap F_v^2|+|F_v^1-F_v^2|\nonumber\\
 \geq&|N_{F_v^1\cap F_v^2}(v)\cup N_{F_v^1\cap F_v^2}(x)|+1\nonumber\\
 \geq&|N_{F_v^1\cap F_v^2}(v)|+|N_{F_v^1\cap F_v^2}(x)|+1\nonumber\\
 \geq&(n-h-2)+(n-3)+1\nonumber\\
 \geq&n-h+1.   \nonumber
\end{align}
This is a contradiction, so there is no fault-free edges between $u$ and $B$ except $uv$.   By lemma \ref{neighbours}, $u$ is not adjacent to $w$, so, ${N_{G'}(u)-v}\subset(F_v^1\cap F_v^2)$.   Therefore, for $n\geq4$,
\begin{align}
n-h\ge |F_v^1|=&|F_v^1\cap F_v^2|+|F_v^1-F_v^2|\nonumber\\
 \geq&|N_{F_v^1\cap F_v^2}(u)\cup N_{F_v^1\cap F_v^2}(v)|+1\nonumber\\
 \geq&|N_{F_v^1\cap F_v^2}(u)|+|N_{F_v^1\cap F_v^2}(v)|+1\nonumber\\
 \geq&(n-1-1)+(n-h-2)+1\nonumber
\end{align}
 a contradiction.

Thus
$t_h^{e}(Q_n)\geq n-h,  $ the Lemma holds.
\hfill\rule{1mm}{2mm}

Combining Lemmas \ref{upper} and \ref{lower}, we have the following theorem.

\begin{thm} Let $h, n$ be positive integers with $n\geq5$ and $1\leq h\leq n-1$,  then  under both the PMC model and  the MM$^*$model $t_h^{e}(Q_n)=n-h$.
\end{thm}
\section{Conclusions}
In this paper, we propose the $h$-edge tolerable diagnosability  as a measure  for the diagnosis capability analysis of multiprocessor systems with both link and node faults. Then
we study the $h$-edge tolerable diagnosability $t_h^{e}(Q_n)$ of the hypercube $Q_n$  under both the PMC and  the MM$^*$model. We prove that $t_h^{e}(Q_n)=n-h$ for  $n\geq5$ and $1\leq h\leq n-1$.   The $h$-edge tolerant diagnosability of other interconnection networks are still to be determined in future.

\bibliographystyle{IEEEtran} \bibliography{IEEEabrv,H-edgediagnosability}
\end{document}